\documentclass[pra]{revtex4-2}

\usepackage{graphicx}
\usepackage{color}

\begin{document}

\title{A possible solution to the which-way problem of quantum interference}

\author{Holger F. Hofmann}
\email{hofmann@hiroshima-u.ac.jp}
\author{Tomonori Matsushita}
\author{Shunichi Kuroki}
\author{Masataka Iinuma}
\affiliation{
Graduate School of Advanced Science and Engineering, Hiroshima University,
Kagamiyama 1-3-1, Higashi Hiroshima 739-8530, Japan
}

\begin{abstract}
It is commonly assumed that the observation of an interference pattern is incompatible with any information about the path taken by a quantum particle. 
Here we show that, contrary to this assumption, the experimentally observable effects of small polarization rotations applied in the slits of a double slit experiment indicate that individual particles passing the slits before their detection in the interference pattern are physically delocalized with regard to their interactions with the local polarization rotations. 
The rate at which the polarization is flipped to the orthogonal state is a direct measure of the fluctuations of the polarization rotation angles experienced by each particle.
Particles detected in the interference maxima experience no fluctuations at all, indicating a presence of exactly one half of the particle in each slit, while particles detected close to the minima experience polarization rotations much larger than the local rotations, indicating a negative presence in one of the slits and a presence of more than one in the other.
\end{abstract}

\maketitle

\section{Introduction}

The mystery of quantum mechanics is often illustrated using the double slit interference of a single quantum particle. Even though the interference pattern is formed only after a large number of particles have been observed, the detection probabilities suggest that each individual particle must have sensed the presence of both slits. This observation appears to be at odds with the seemingly natural assumption that each particle can only pass through one slit or the other. The original explanation of the problem focused on the complementarity of interference and which-way measurements, resulting in a number of experiments aimed at observing a particle as it moves through the slits \cite{Scu91,Dur98,Sch99,Sch02,Bar09}. The main conclusion drawn from this experimental approach to the which-way problem was that which-way information is incompatible with the observation of an interference pattern, resulting in a quantitative trade-off of the visibilities \cite{Eng96}. However, this approach to the which-way problem assumed that the evidence for a particle position must be obtained in the form of a single yes/no answer to the question of whether the particle passed through a particular slit or not. This may seem to be a reasonable assumption since a detection of the particles in the slits would always produce a yes/no result of this kind. However, quantum theory has shown that the assignment of hidden realities corresponding to the outcomes of measurements that have not been actually performed can result in unresolvable inconsistencies \cite{KS,FR,Bru18}. 
It is therefore unlikely that the which-path problem can be resolved by claiming that the path of the particles is merely unknown and could be described by non-contextual hidden variables. Instead, we should consider the possibility that the localization of a particle depends on the measurement context. In particular, it should not be ruled out that individual particles might be physically delocalized between the slits when they are detected as part of an interference pattern.

Quantum theory itself describes a consistent assignment of measurement dependent values different from the eigenvalues of an observable in terms of the weak values defined by an initial state and a specific measurement outcome \cite{Tol07,Dre10,Pus14,Pia16}. The reason why the observation of weak values has not resolved the controversies concerning the measurement dependence of physical properties is that the available evidence is obtained by averaging over very noisy data, making it difficult to evaluate the physical property of an individual system with sufficient precision. To avoid any controversy regarding the meaning of the statistics, it has been suggested that the qualitative which-way information obtained from weak effects is sufficient to obtain non-trivial path information \cite{Vai13,Vai14} and corresponding experiments have been performed to confirm this prediction \cite{Dan13,Gep18,Spo19,Qi20,Daj21}. 
However, the results have merely highlighted the difficulty of assigning specific paths to individual particles using only qualitative statistical evidence \cite{Gri16,Eng17,Pel19,Vai20,Cor21,Han21}.

In order to genuinely resolve the which-way problem, it is necessary to carefully consider the experimental evidence obtained through weak interactions. In particular, we need to consider the difference between statistical averages and the physical effects experienced by individual particles. In this regard, it is important to note that weak interactions are fully coherent, resulting in a precise definition of the effects of an observable on a quantum mechanical probe system \cite{Dzi19}. This observation strongly suggests that weak values correctly describe the effects of individual systems on their environment. However, it is somewhat difficult to confirm this observation in an experiment. The approach proposed in \cite{Dzi19} requires a superposition of a probe that interacted with the system and a probe that did not interact with the system. This method is certainly sufficient to confirm the validity of the theoretical analysis, but it may not be sufficient to convince sceptics who suspect that quantum superpositions merely hide hidden statistics. This is particularly important with respect to the double-slit problem, where the established wisdom seems to suggest that the position in the interference pattern provides no information about the paths, so that the uncertainty of the path after the detection of a particle in the interference pattern should be the same as the uncertainty of the path in the initial state. It is therefore important to ask whether it might be possible to observe experimental evidence of the fluctuations of the paths taken by individual particles in an interference experiment. Such an experimental method of evaluating the conditional path statistics associated with specific positions of a particle in an interference measurement could resolve the question of whether the particle was physically delocalized as it passed through the slits when it is subsequently observed in an interference pattern, even if it remains unclear which of the two paths each particle preferred. As shown in a recent letter, it is indeed possible to use a weak interaction with a quantum probe to experimentally evaluate the fluctuations of the physical property that acts on the probe for a specific final measurement outcome \cite{Hof21}. If weak values describe the effects of individual systems, they can be compensated by a feedback acting on the probe. Any difference between the weak value used to define the feedback and the actual value for the individual system then shows up in the experimental data as a rate of flips in the two-level quantum probe. This method has already been used to demonstrate the physical delocalization of individual particles between the arms of a two-path interferometer when the paths are biased and the phase difference between the paths is zero \cite{Lem22}. Here we apply the same method to the original which-way problem of the double slit experiment, where the symmetry of the paths means that there is no preferred path of the particles. In this situation, the average presence of the particles in the slits is one half for all particles detected in the interference pattern and a weak measurement would merely confirm this obvious fact. However, it is unclear whether this average presence is obtained because the particles pass through one of the two slits with a probability of 1/2 each as observed in a which-way measurement, or if each particle is physically distributed between the slits. To distinguish these possibilities, we apply opposite polarization rotations in the two slits and consider their effect on the particles detected in the interference pattern. If the particles passed through only one slit they would experience a polarization rotation of the same magnitude in each of the slits. We can therefore predict the rate at which the polarization would have to flip to the opposite state if the particle had been localized in only one of the slits. If each particle is physically delocalized, the polarization rotation experienced by each particle is given by the average of the rotation angles applied in each slit, resulting in a perfect cancellation of the rotation. The rate at which the polarization flips thus drops to zero when the particle is physically delocalized with a presence of 1/2 in each of the slits. 

The results of our analysis show that the rate of polarization flips depends on the position in the interference pattern at which the particle is detected. A perfect physical delocalization with no polarization flips is only observed at the interference maximum. However, the rate of flips remains lower than the rate required by localization in the slits as long as the interference is constructive. Interestingly, particles detected at the turning points between constructive and destructive interference show the exact rate of polarization flips required by a localization of the particle in one of the slits, indicating that even the accidental absence of interference is associated with a localization of the particles in the slits. Finally, the rate of polarization flips exceeds the value required by localization when the interference is destructive. This result may seem odd, but it is consistent with the assignment of a negative presence in one slit and a presence greater than one in the other, where the negative presence indicates that the particle experiences a polarization rotation in the direction opposite to the one applied in the slit. The observed fluctuations of the polarization rotations experienced by the particles therefore suggest that the delocalization of the particles in destructive interference can also involve a negative presence of each particle in one of the slits. 

We would like to emphasize that the discussion of particle delocalization based on the experimental evidence obtained from small polarization rotations reveals new and unexpected aspects of the which-way problem. The possibility of identifying the delocalization of individual particles observed in an interference pattern changes the physical meaning of the concept of delocalization. In much of the previous literature, ``delocalization'' merely refers to the abstract mathematical concept of superpositions (see e.g. \cite{Dup22}). Here we show that such a formal definition of ``delocalization'' can be highly misleading. Hopefully our work will help to replace such unclear and confusing concepts with a more specific operational definition based on the physical response of particles to locally applied forces and effects.

\section{Weak entanglement between paths and polarization}

The experimental setup we are considering is very similar to the previous setups used to confirm the uncertainty relations and their relation with quantum entanglement \cite{Scu91,Dur98,Sch99,Sch02,Bar09}. In this kind of setup, the path of a photon through the slits can be marked by polarization rotations applied in the slits. We now use this procedure to introduce a specific operational definition of what we mean by the path of a particle. Since the rotations of polarization are applied locally in only one specific path each, the presence of a particle in the path is defined by the relation between the polarization rotation experienced by the particle and the rotation applied in the path. A particle has a presence of one in the path if it experiences exactly the rotation associated with the path and a presence of zero if it does not experience that rotation at all. Importantly, we can use this definition to distinguish between a random selection of paths and a particle that is physically distributed between different paths because the polarization statistics of a random rotation can be distinguished from a polarization rotated by a precise fraction of the angle associated with a specific path. 

For convenience, we will consider the application of rotation angles of opposite sign to the two slits. If all photons pass through only one of the two slits, their polarizations will always rotate, either by $\phi_1=+\theta$ or by $\phi_2=-\theta$ and the outcome polarizations can be explained in terms of the corresponding rotation statistics. Using $\mid 1 \rangle$ and $\mid 2 \rangle$ for the paths through slit 1 and the path through slit 2, respectively, $\mid H \rangle$ for the initial horizontal polarization of the input photons and $\mid V \rangle$ for the vertical polarization component, the state after the polarization rotations reads
\begin{equation}
\label{eq:state}
\mid \psi \rangle = \frac{1}{\sqrt{2}} \left( \left(\cos(\theta)\mid H \rangle + \sin(\theta) \mid V \rangle \right) \otimes \mid 1 \rangle + \left(\cos(\theta)\mid H \rangle - \sin(\theta) \mid V \rangle \right)\otimes \mid 2 \rangle \right).
\end{equation}
Note that the direction of the polarization rotation is encoded in the sign of the coherence between the polarization components $\mid H \rangle$ and $\mid V \rangle$. In an interference experiment, this coherence is transferred to the screen resulting in a correlation between the position on the screen and the polarization. 
When the distance to the screen is much larger than the distance between the slits, the position at which the photon is detected on the screen depends only on the transverse momentum $p$ of the particle, which is conserved in free space propagation. It is therefore convenient to identify the eigenstates of positions on the screen with the momentum eigenstates $\mid p \rangle$. Since each value of $p$ corresponds to a well defined spatial position on the screen, we will express the local detection of the particle far away from the screen using the momentum $p$ to denote the position of the detection on the screen.
The contributions of the two slits separated by a distance of $d$ can then be given by
\begin{eqnarray}
\langle p \mid 1 \rangle = f(p) \exp\left(+ i \frac{d}{2 \hbar} p \right),
\nonumber \\
\langle p \mid 2 \rangle = f(p) \exp\left(- i \frac{d}{2 \hbar} p \right),
\end{eqnarray}
where $f(p)$ represents the sinc-function of single slit diffraction. 
The interference pattern on the screen can be obtained from the state in Eq.(\ref{eq:state}) by summing over the contributions from the orthogonal polarizations,
\begin{equation}
\label{eq:screen}
|\langle p,H \mid \psi \rangle|^2 + |\langle p,V \mid \psi \rangle|^2 
= (f(p))^2 \left(1 + \cos(2 \theta)\cos\left(\frac{d}{\hbar} p \right)\right).
\end{equation}
For small polarization rotations $\theta$ the reduction in visibility is negligibly small ($\cos (2\theta) \approx 1$). Nonetheless the state in Eq.(\ref{eq:state}) describes entanglement between the path and the polarization of the photon, making it possible to extract information about the path taken by the photon from measurements of polarization performed on photons detected as part of the interference pattern. The important challenge is to identify the effects of the polarization rotation associated with the paths on the polarization of the photon. 

\section{Correlations between path uncertainties and position in the interference pattern}

It is necessary to carefully consider the effects of the polarization rotations applied in the slits on the polarization of the particles detected in the interference pattern. The fact that this effect depends strongly on the position $p$ in the interference pattern at which the particle is detected strongly suggests that the paths taken by the particle depends on the position $p$ and the associated type of interference. This dependence can be expressed by
\begin{equation}
\label{eq:pattern}
\langle p \mid \psi \rangle = \sqrt{2} f(p) \left( \cos(\theta) \cos\left(\frac{d}{2 \hbar} p \right) \mid H \rangle + i \sin(\theta) \sin\left(\frac{d}{2 \hbar} p \right) \mid V \rangle\right).
\end{equation}
The interference effect has converted the rotation of polarization in the slits into an elliptical polarization with an unchanged orientation of the main axes along the horizontal ($H$) and vertical ($V$) directions. This means that there is no preferred rotation direction. The best estimate for the rotation angle is $\phi=0$, corresponding to the average of the local rotations of $\phi_1=+\theta$ and $\phi_2=-\theta$. However, a statistical mixture of these two rotations would result in a polarization density matrix of 
\begin{equation}
\hat{\rho}_{\mathrm{pol.}} = \cos(\theta)^2 \mid H \rangle \langle H \mid + \sin(\theta)^2 \mid V \rangle \langle V \mid.
\end{equation}
We can therefore define the uncertainty $\delta \phi$ of the polarization rotation angle in terms of the probability of finding a vertically polarized photon. For sufficiently small probabilities $P(V)$, 
\begin{equation}
\delta \phi^2 \approx P(V).
\end{equation}
According to Eq.(\ref{eq:pattern}) the uncertainty of the polarization rotation angle depends on the position $p$ within the interference pattern. The conditional probabilities of vertical polarization are given by
\begin{equation}
P(V|p) = \frac{\tan(\theta)^2 \tan\left(\frac{d}{2 \hbar} p \right)^2}{1+\tan(\theta)^2 \tan\left(\frac{d}{2 \hbar} p \right)^2}.
\end{equation}
For sufficiently small values of $\theta$, this conditional probability corresponds to a rotation angle uncertainty of 
\begin{equation}
\delta \phi^2(p) \approx \theta^2 \tan\left(\frac{d}{2 \hbar} p \right)^2.
\end{equation}
Since we know that the fluctuations in the polarization rotations are caused by the difference between the rotation angles $\theta$ and $-\theta$ in the slits 1 and 2, the uncertainty conditioned by the position $p$ in the interference pattern corresponds directly to an uncertainty in the path of the particle. Significantly, this uncertainty is different from the uncertainty expected for a particle localized in one of the two slits. The dependence of the fluctuations of the polarization angle on the position $p$ in the interference pattern therefore suggest that 
each individual particle is physically delocalized between the slits, permitting it to experience a proportional combination of the polarization rotations $\theta$ and $-\theta$ applied locally in each slit.

\begin{figure}[th]
\begin{picture}(340,250)
\put(0,0){\makebox(340,180){
\scalebox{0.8}[0.8]{
\includegraphics{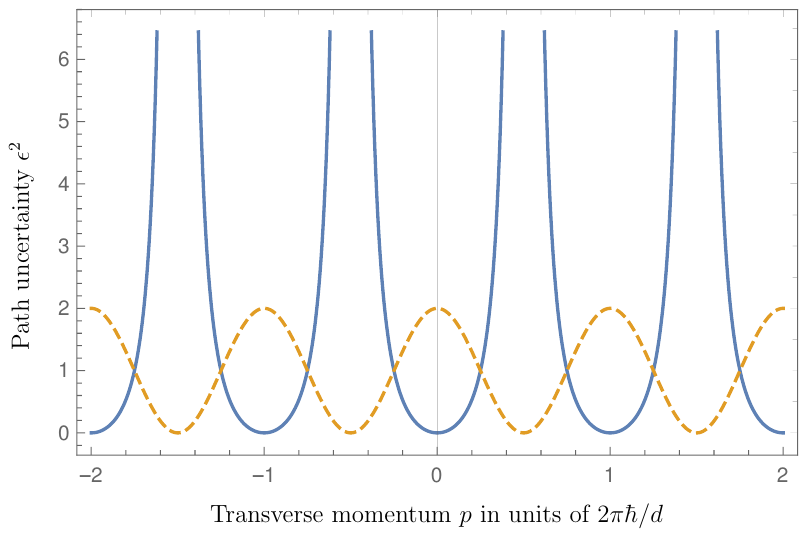}}}}
\end{picture}
\caption{\label{fig1}
Dependence of path uncertainty on the position at which the particle arrives in an interference experiment. The dotted line shows the interference effect normalized to an average value of one. Path uncertainties smaller than one are obtained for constructive interference while path uncertainties larger than one are obtained for destructive interference.  
}
\end{figure}

For a more detailed analysis it is convenient to convert the fluctuations of polarization rotations into a normalized measure of path fluctuations,
\begin{equation}
\epsilon^2(p) = \frac{\delta \phi^2(p)}{\theta^2}.
\end{equation}
This path fluctuation is independent of $\theta$, indicating that it is not an artefact caused by the polarization rotations. 
Instead, the polarization rotations provide a reliable quantitative measure of the conditional path uncertainty, where the eigenvalues assigned to the paths are $+1$ and $-1$. If the paths are completely unknown, the value of $\epsilon^2(p)$ should always be exactly one. However, we find that completely different values are obtained for different positions $p$ in the interference pattern. For sufficiently small rotation angles $\theta$, the interference pattern is undisturbed and the probability density is
\begin{equation}
\rho_{\mathrm{screen}}(p) =  |f(p)|^2 \left(1+\cos\left(\frac{d}{\hbar} p \right)\right).
\end{equation}
The path fluctuations depend only on the cosine function that distinguishes the minima and the maxima of the interference pattern from each other,
\begin{equation}
\epsilon^2(p) = \frac{1-\cos\left(\frac{d}{\hbar} p \right)}{1+\cos\left(\frac{d}{\hbar} p \right)}
\end{equation}
Fig. \ref{fig1} shows the dependence of the path fluctuations on the position in the interference pattern. In the maxima of the interference pattern ($\cos(d p /\hbar)=+1$), the path fluctuation is zero. This means that the rotation of $+\theta$ and $-\theta$ in the two slits cancel each other perfectly for each photon that passes through the slits, indicating that photons detected in the maxima are equally split between the paths. For constructive interferences ($0<\cos(d p / \hbar)<1$) the fluctuations are smaller than one, indicating that a non-vanishing fraction of the particle is present in each slit. For destructive interferences ($-1<\cos(d p /\hbar)<0$) the fluctuations exceed 1, indicating that the proportional contribution of one slit must be negative to allow for a contribution greater than $\theta$ in the other slit. In the interference minima, the fluctuation diverges to infinity, representing the maximal change of polarization from $H$ to $V$ at these points. The actual result obtained at a minimum is of course finite, but this can be explained by the changes in the detection probability from an original value of zero to a finite value. For any specific value of $p$ with non-vanishing detection probability at $\theta=0$, the polarization rotations can always be made sufficiently small so that the changes to the probabilities of finding the photon at $p$ are minimal and the theoretically predicted value of $\epsilon^2(p)$ can be obtained from the values of $P(V|p)\ll 1$. The result therefore indicates that a tiny polarization rotation in the slits will cause a much larger polarization change at values of $p$ close to the minimum. 

\section{Implications for our fundamental understanding of uncertainties}

It is generally assumed that the uncertainty principle prevents us from obtaining any meaningful which-path information during an interference experiment. However, our results indicate that it is possible to identify the conditional uncertainties of a physical property without any direct measurement of the path taken by the particle. As shown in \cite{Hof21}, the fluctuations of the path observable given by the operator $|1><1|-|2><2|$ can be observed in the randomization of the polarization direction caused by the quantum controlled polarization rotations. Although it is tempting to misinterpret this operation as a random mix of rotations by $+\theta$ in path 1 and by $-\theta$ in path 2, the results show that the quantitative measure of the path presence  described by the operator $|1><1|-|2><2|$ fluctuates in a manner that is not consistent with this classical interpretation. The evidence clearly shows that we need to abandon the idea that each particle must be located in one of the slits. Instead, the quantitative evaluation of the changes to the polarization conclusively demonstrates that each particle experiences a specific fraction of the spin rotations in the two slits. In particular, a rotation of $\delta \phi>\theta$ indicates a negative presence of the particle in one path and a corresponding presence larger than one in the other. Obviously this result is fundamentally different from any direct detection of the particle in one of the slits. It is therefore important to remember that it is impossible to identify a small rotation angle by directly measuring the output polarization. There is no practical contradiction between the precise assignment of rotation angles of $+\theta$ and $-\theta$ in a which-way measurement and the assignment of fluctuations $\epsilon^2(p)$ larger and smaller than one to the different particle position $p$ in the interference pattern. Both assignments are consistent with the polarization statistics of the photons after the rotations have been applied. The information about the path taken by the photon conveyed by the path uncertainty $\epsilon^2(p)$ is provided exclusively by the measurement outcome $p$ of the interference measurement. The effects of local polarization rotations merely confirm that there is a correlation between the presence of each particle in the paths and the measurement outcome $p$ of an interference measurement. Within the framework of the theory, this peculiar dependence of path presence on the outcomes $p$ is possible because 
the entanglement between polarization and paths in Eq.(\ref{eq:state}) expresses a non-trivial correlation between the outcomes of the interference measurements and the polarization of the photons. Since free space propagation does not change the polarization, the polarization selected by the detection at position $p$ on the screen must originate from the local polarization rotations at the slits. Effectively, the entangled state expresses the conditional polarization rotation obtained at $p$ as a superposition of the two polarization rotations applied at the slits. It has been pointed out in previous works that such interferences between two different unitary operations can result in a transformation that seems to be qualitatively different from the two operations that interfered with each other \cite{Yok16,Hollis}. In the present interference experiment, the total change of polarization observed at each point in the interference pattern cannot be explained in terms of a statistical distribution of the local rotation angles.
Instead, the effects of the local rotations are conditioned by the quantum interference in the final measurement, even though the local rotations are applied before the particles are detected in the interference pattern. 

The dependence of path fluctuations on the position $p$ at which the particle is found in the interference pattern is directly observable in the form of the conditional probabilities $P(V|p)$ describing the magnitude of the effect of the local polarization rotations. This evidence demonstrates conclusively that the outcomes of interference measurements provide non-trivial which-path information. In the present case, the paths are symmetric and the information is obtained by evaluating the fluctuations. 
This is a special case of the general method presented in \cite{Hof21}, where no feedback compensation is needed because the best estimate of the differences between the paths is always zero, independent of the measurement outcome. In the case of two-path interference at zero phase difference, it has already been shown that error-free fractional values of the path presence can be obtained when the input state is biased in the two paths \cite{Lem22}.
In general, the precise analysis of the effects of weak operations in the paths of an interference experiment indicates that there is a well-defined correlation between the outcome of the interference experiment and the presence of each individual particle in the paths. 
We would like to stress that the only empirically valid definition of the presence of a particle at a specific location is given by the effects of local interactions with the particle. In the present paper, we have solved the which-way problem by evaluating the effects of local polarization rotations on the polarization of particles observed in the interference pattern. The delocalization of the particles that we observe is physical because it has quantitative physical effects. It may seem counter intuitive that the particles would be localized if they were detected in the slits, but it should be kept in mind that our intuition is based on the assumption that we can observe objects with negligible interactions. In the quantum mechanical limit, weak interactions provide statistical information that is fundamentally different from the direct observation of reality that our intuition imagines. As we have shown here, the fluctuations of weak interactions are nonetheless sufficient to identify the values taken by the particle presence between the source and the detector. 

\section{Conclusions}

It should not come as such a great surprise that the which-way reality depends on the type of final measurement that is actually performed. The Kochen-Specker theorem and the closely related consistency paradoxes have already demonstrated that physical properties necessarily depend on the measurement context \cite{KS,FR,Bru18}. In this sense the present analysis simply completes the picture by providing direct evidence that the which-way problem can be resolved by a physical delocalization of the particles in the context of double slit interferences. Specifically, we have shown that this delocalization of particles between the slits has experimentally observable consequences that allow us to determine the precise nature of the physical delocalization of particles in an interference experiment. We find that the delocalization depends on whether the interference is destructive or constructive, indicating that we have discovered a fundamental aspect of causality in quantum mechanics that has been overlooked in previous discussions of the which-way problem. The reason why this fundamental feature of the theory has been overlooked for so long might be that quantum mechanics is usually represented as random and therefore acausal. However, a closer look at the formalism suggests that quantum interferences determine the physics of processes in far more detail than the randomness of the measurement outcomes suggests. It may therefore be time to reconsider the physics described by the quantum formalism in the light of the evidence presented above.

\vspace{0.5cm}

\section*{Statements}
On behalf of all authors, the corresponding author states that there is no conflict of interest. 
The present paper has no associated data.

\end{document}